%% file: main.tex
\documentclass[conference]{IEEEtran}
\IEEEoverridecommandlockouts
\usepackage{cite}
\usepackage{amsmath,amssymb,amsfonts}
\usepackage{graphicx}
\usepackage{textcomp}
\def\BibTeX{{\rm B\kern-.05em{\sc i\kern-.025em b}\kern-.08em
    T\kern-.1667em\lower.7ex\hbox{E}\kern-.125emX}}
\usepackage{comment}
\usepackage{bm}

\usepackage[utf8]{inputenc} 
\usepackage[T1]{fontenc}    
\usepackage{url}            
\usepackage{booktabs}       
\usepackage{diagbox}        
\usepackage{makecell}
\usepackage{nicefrac}       
\usepackage{microtype}      
\usepackage{xspace}
\usepackage{bm}
\usepackage{times}
\usepackage[ruled,vlined]{algorithm2e}
\usepackage{xcolor}
\definecolor{customblue}{rgb}{0.36, 0.55, 0.75}
\usepackage[colorlinks=true,linkcolor=magenta, urlcolor=magenta,citecolor=customblue, anchorcolor=magenta]{hyperref}
\usepackage{marvosym}
\usepackage{booktabs}
\usepackage{tabularx}
\usepackage{multirow}

\usepackage{wrapfig}

\usepackage{bbm}
\usepackage{multirow}
\usepackage{float}
\usepackage{subcaption}
\usepackage{scalerel}

\usepackage{array}
\newcolumntype{H}{>{\setbox0=\hbox\bgroup}c<{\egroup}@{}}

\newcommand{\cut}[1]{}  
\newcommand{\std}[1]{\scriptsize{$\pm$#1}}

\def\BibTeX{{\rm B\kern-.05em{\sc i\kern-.025em b}\kern-.08em
    T\kern-.1667em\lower.7ex\hbox{E}\kern-.125emX}}
\begin{document}

\title{What Causes Performance Degradation in Cross-Subject EEG Classification?
}

\author{
Yihe Wang,$^*$
Taida Li,$^*$
Yujun Yan,
Wenzhan Song,
Xiang Zhang\textsuperscript{\Letter} 
\thanks{Yihe Wang, Taida Li, and Xiang Zhang are with the Department of Computer Science,  University of North Carolina-Charlotte, United States. E-mail: ywang145@charlotte.edu, tli14@charlotte.edu, xiang.zhang@charlotte.edu.}
\thanks{Yujun Yan is with the Department of Computer Science, Dartmouth College, United States. E-mail: yujun.yan@dartmouth.edu}
\thanks{Wenzhan Song is with the School of Electrical and Computer Engineering, University of Georgia, United States. E-mail: wsong@uga.edu}
}

\maketitle
\def\thefootnote{*}\footnotetext{Equal Contribution.}

\begin{abstract}
\input{000abstract.tex}
\end{abstract}


\section{Introduction}
\label{sec:intro}

\input{010intro}

\section{Preliminaries and Datasets}
\label{sec:preliminary}

\input{020preliminaries}

\section{Experiments and Results}
\label{sec:experiment}

\input{030experiments}

\section{Discussion}
\label{sec:discussion}

\input{040discussion}

\section{Conclusion}
\label{sec:conclusion}

\input{050conclusion}

\bibliographystyle{unsrt}  
\bibliography{refs}

\end{document}

%% file: 000abstract.tex
Cross-subject Electroencephalography (EEG) classification typically achieves significantly lower performance than subject-dependent settings. Although this phenomenon has been widely observed in the literature, the underlying causes have not been systematically studied. In this paper, we design a series of controlled experiments to investigate the mechanisms behind the performance drop in cross-subject EEG classification across different EEG tasks. We show that the performance degradation can generally be attributed to two factors: inter-subject variability and shortcut learning. Specifically, multi-class-per-subject EEG classification tasks, such as motor imagery, emotion recognition, and ERP stimulus classification, are mainly affected by inter-subject variability, whereas single-class-per-subject EEG classification tasks, such as brain disease detection, are primarily influenced by shortcut learning based on subject-specific features. These findings provide new insights into the challenges of cross-subject EEG classification and emphasize the importance of appropriate evaluation protocols in EEG research. The code is available at \url{https://github.com/DL4mHealth/EEG-Cross-Subject}.

%% file: 010intro.tex
Cross-subject EEG classification remains a major challenge in EEG-based brain-computer interfaces and clinical EEG analysis. In many real-world applications, models are expected to generalize to previously unseen subjects rather than subjects seen during training~\cite{pandey2022subject,zheng2025fapex}. However, cross-subject EEG classification generally yields lower performance than subject-dependent settings, in which the training and test sets may contain samples from the same subjects. Although this performance gap has been widely reported and discussed in prior studies~\cite{kunjan2021necessity,yu2026decentralized,hu2025channel}, its underlying causes remain poorly understood.

\begin{figure}  
    \includegraphics[width=1.0\columnwidth]{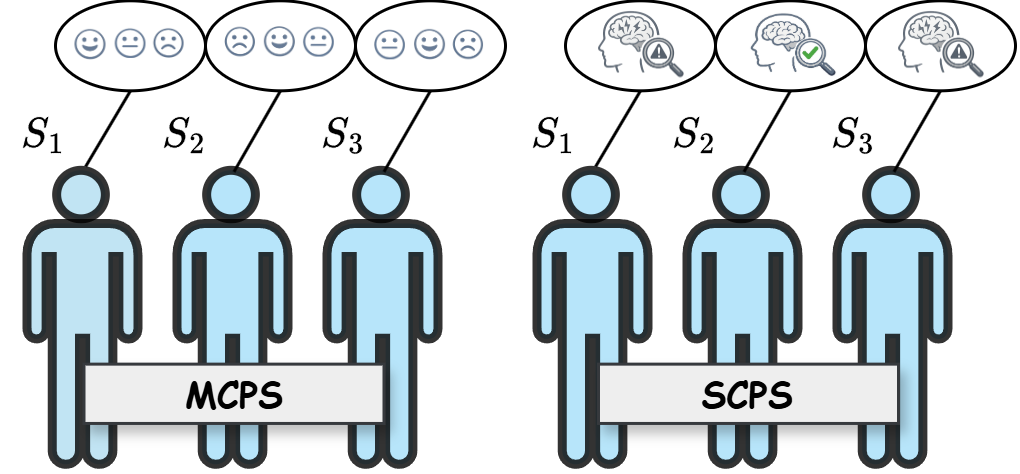}
    \caption{\textbf{Two types of EEG task.} 
    In MCPS tasks, labels are assigned to samples and may vary across samples within a subject. In SCPS tasks, each subject is assigned a single label, and all samples from that subject share the same label.
    }
    \label{fig:eeg_task}
    \vspace{-3mm}
\end{figure}

EEG classification tasks can generally be divided into two categories: \textbf{multi-class-per-subject (MCPS)} tasks and \textbf{single-class-per-subject (SCPS)}. Figure~\ref{fig:eeg_task} illustrates the distinction between these two task types. In MCPS tasks, a single subject transitions across multiple class states over time. Typical examples include motor imagery~\cite{vafaei2025transformers,hameed2025enhancing}, emotion recognition~\cite{li2022eeg,gkintoni2025neural}, seizure detection~\cite{zheng2025fapex,li2025cnn}, and ERP stimulus classification~\cite{bagchi2022eeg}. In SCPS tasks, each subject is assigned a fixed label that does not change across time. Representative examples include many brain disorder detection tasks, such as Alzheimer’s disease (AD) detection~\cite{watanabe2024deep,wang2025lead}, Parkinson’s disease (PD) detection~\cite{maitin2022survey}, and depression detection~\cite{elnaggar2025depression}. Although cross-subject performance degrades in both task types, the underlying causes of this decline, as well as the relative extent of degradation between the two tasks, remain open questions that have not yet been systematically investigated.

Most existing works attribute the performance drop in cross-subject EEG classification to inter-subject variability, i.e., differences in EEG signal distributions across subjects~\cite{gkenios2022diagnosis,wang2024medformer}. These differences may arise from anatomical differences, electrode placement, brain activity patterns, recording environments, and other factors. As a result, models trained on some subjects may not generalize well to unseen subjects. Although this explanation holds for most MCPS scenarios, inter-subject variability alone is insufficient to fully account for the severe performance degradation observed on SCPS tasks.

In this paper, we propose that the cross-subject performance gap stems from two primary mechanisms. First, inherent inter-subject variability creates a domain shift, preventing models from generalizing learned feature distributions to unseen subjects. Second, in certain cases, models are highly susceptible to shortcut learning and may rely on subject-specific features rather than task-relevant information. To investigate these two factors, we design a series of controlled experiments across multiple datasets and tasks to analyze the causes of performance degradation in cross-subject EEG classification. Our experiments show that \textbf{MCPS EEG classification tasks are mainly affected by inter-subject variability, whereas SCPS EEG classification tasks are largely influenced by shortcut learning based on subject-specific features.} We hope these findings provide a new perspective on cross-subject EEG classification, emphasize the importance of proper evaluation protocols, and inspire future research on method design to address these challenges in EEG analysis.

%% file: 020preliminaries.tex
\subsection{Subject-Dependent and Subject-Independent Setup}
\label{sub:sub_dep_indep}
\begin{figure}  
    \includegraphics[width=1.0\columnwidth]{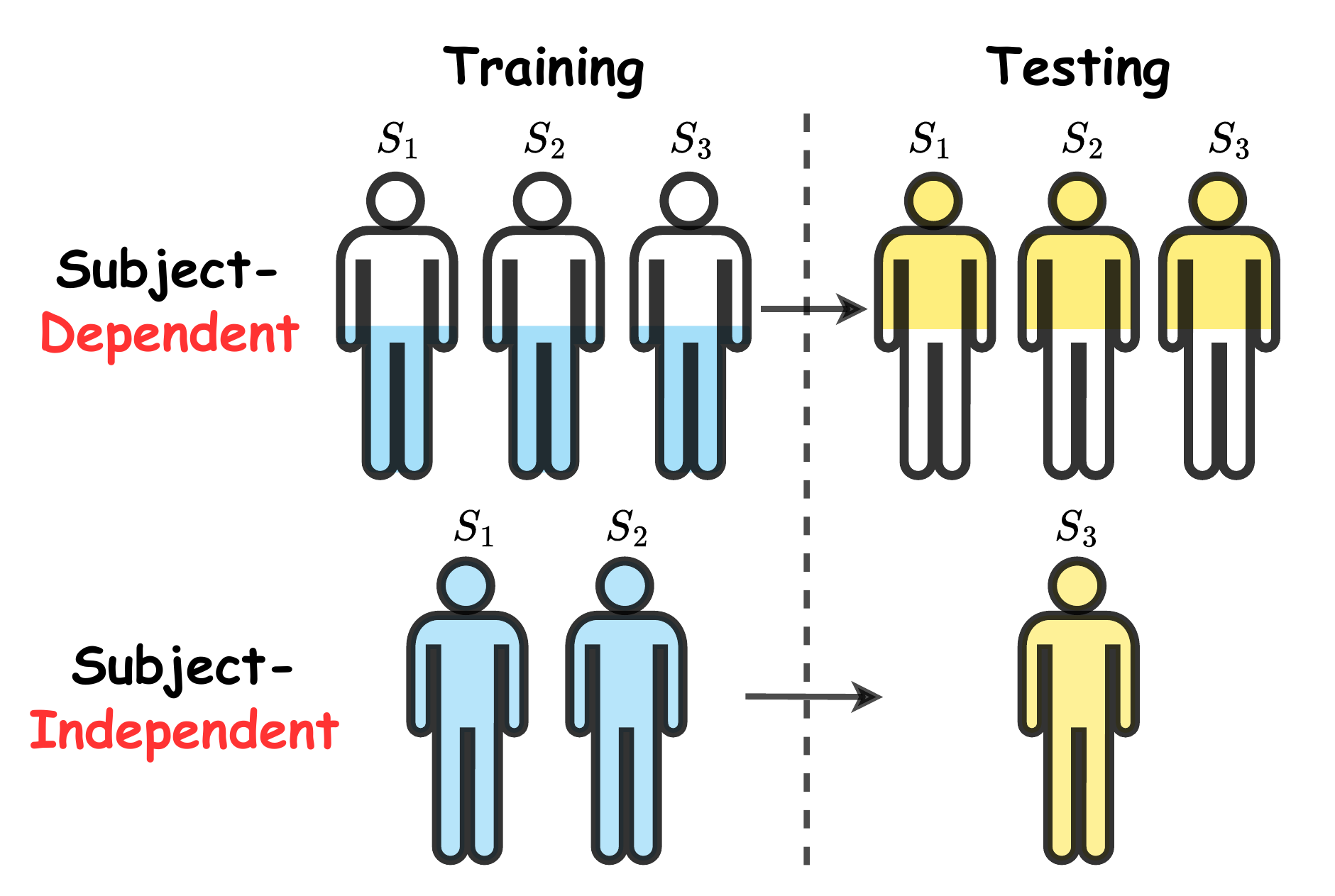}
    \caption{\textbf{Subject-dependent/independent setups.} 
    In the subject-dependent setup, samples from the same subject may appear in both the training and test sets, while in the subject-independent setup, samples from the same subject are exclusively in either the training or test set. (Adopted from~\cite{wang2024contrast})
    }
    \label{fig:subject-dependent-independent}
    \vspace{-3mm}
\end{figure}

\textbf{Subject-Dependent:} In the subject-dependent setup, subject identity is ignored during data splitting. All samples from different subjects are pooled together and randomly divided into training and test sets. Consequently, samples from the same subject may appear in both training and test sets~\cite{sharma2025graph,acharya2025eegconvnext}. \textbf{Subject-Independent:} In contrast, the subject-independent setup performs cross-subject evaluation and better reflects real-world deployment scenarios. In this setup, the dataset is split by subject rather than by individual samples, ensuring that all samples from a given subject are assigned exclusively to either the training or test set~\cite{albuquerque2019cross,wang2024contrast}. Let $S_{train}$ and $S_{test}$ denote the sets of subjects appearing in the training and test sets. In the subject-dependent setup, the subject sets are not disjoint: $S_{train} \cap S_{test} \neq \emptyset$, whereas in the subject-independent setup, the subject sets are disjoint: $S_{train} \cap S_{test} = \emptyset$. Figure~\ref{fig:subject-dependent-independent} illustrates the difference between these two setups.

\input{tables/datasets}

\subsection{Datasets}
\label{sub:datasets}
We use five datasets in this study: mTBI-ODD~\cite{cavanagh2019erps}, MMIDB~\cite{schalk2004bci2000} ADFTD~\cite{miltiadous2023dataset}, PD-RS~\cite{singh2023evoked}, and AOPD~\cite{cavanagh2021eeg}. The mTBI-ODD dataset is used for an MCPS EEG classification task to classify ERP stimulus types into standard, target, and novel oddball conditions. The MMIDB dataset is used in an MCPS EEG classification task to decode motor imagery of opening and closing both fists or both feet. The ADFTD dataset is used in an SCPS EEG classification task to distinguish healthy controls (HC) from patients with AD and Frontotemporal Dementia (FTD). The PD-RS dataset is used in an SCPS EEG classification task to detect PD from HC subjects. The AOPD dataset contains both MCPS and SCPS labels: the sample-level labels indicate whether the stimulus is oddball (standard vs. deviant), while the subject-level labels indicate whether the subject is healthy or has PD. The preprocessing procedures for ADFTD and PD-RS follow Medformer~\cite{wang2024medformer}, while the preprocessing procedures for mTBI-ODD and AOPD follow the ERP benchmark study~\cite{wang2026benchmarking}, and MMIDB follows the examples provided in the Braindecode library~\cite{braindecode}. The statistics of the processed datasets are summarized in Table~\ref{tab:processed_data}.

%% file: tables/datasets.tex
\begin{table*}[]
    \centering
    \scriptsize
    \caption{\textbf{Dataset Statistics.} The table shows the number of subjects, samples, channels, sampling rate, sample timestamps, EEG classification task types, and labels. 
    }
    \vspace{-2mm}
    \label{tab:processed_data}
    \resizebox{\textwidth}{!}{%
    \begin{tabular}{@{}ll|ccccccc@{}}
    \toprule
    \multicolumn{2}{l|}{\textbf{Datasets}} & \textbf{\#-Subject} & \textbf{\#-Sample} & \textbf{\#-Channel} & \textbf{\#-Timestamps} & \textbf{Sampling Rate} & \textbf{EEG Task Types} & \textbf{Label}  \\ \midrule
    \multicolumn{2}{l|}{\textbf{mTBI-ODD}~\cite{cavanagh2019erps}} &  96  & 24,885  & 61 & 200 &  200Hz  & MCPS (3-class) & Standard vs Target vs Novel \\
    \multicolumn{2}{l|}{\textbf{MMIDB}~\cite{schalk2004bci2000}} &  109  & 4,897  & 64 & 640 &  160Hz  & MCPS (2-class) & Both Fists vs Both Feet \\
    \multicolumn{2}{l|}{\textbf{ADFTD}~\cite{miltiadous2023dataset}} &  88  & 69,752  & 19 & 200 &  200Hz  & SCPS (3-class) & HC vs AD vs FTD \\
    \multicolumn{2}{l|}{\textbf{PD-RS}~\cite{singh2023evoked}} &  149  & 23,839  & 60 & 200 &  200Hz  & SCPS (2-class) & HC vs PD\\
    \multicolumn{2}{l|}{\textbf{AOPD}~\cite{cavanagh2021eeg}} &  50  & 9,830  & 59 & 200 &  200Hz  & \makecell{MCPS (2-class) \\ \& SCPS (2-class)} & \makecell{Standard vs Deviant \\ \& HC vs PD} \\
    \bottomrule
    \end{tabular}
    } 
    \vspace{-2mm}
\end{table*}

%% file: 030experiments.tex
In this section, we conduct a series of experiments to analyze the causes of performance degradation in cross-subject EEG classification. First, we evaluate both MCPS and SCPS EEG tasks under subject-dependent and subject-independent settings to observe the performance drop. Next, to rule out the possibility that the observed differences are driven by dataset-specific factors rather than task formulation, we compare MCPS and SCPS tasks within the same dataset. We analyze performance degradation on a dataset that contains both MCPS and SCPS EEG tasks. We then demonstrate the existence of subject-specific features by conducting subject discrimination experiments. Finally, we conduct a label-permutation experiment to show that the causes of performance degradation in cross-subject classification differ between the MCPS and SCPS EEG tasks. To eliminate performance variance caused by different model architectures, all experiments are conducted using a widely used EEG model, EEGConformer~\cite{song2022eeg}. The datasets are randomly split into 60\%, 20\%, and 20\% for training, validation, and test sets, respectively, either by samples or by subjects, depending on whether a subject-dependent or independent setup is used. Results are reported as the F1 score and AUROC. All experiments are repeated with random seeds 41-45, and the mean and standard deviation are reported.

\input{tables/subject-dep-indep}

\subsection{Performance Drop under Subject-Independent Evaluation}
\label{sub:sub_indep_dep}
We first examine whether model performance decreases, and to what extent, when moving from subject-dependent to subject-independent evaluation on both MCPS and SCPS EEG tasks. We conduct experiments on mTBI-ODD, MMIDB, ADFTD, and PD-RS datasets. The results are presented in Table~\ref{tab:sub_dep_indep}.

The results show that performance decreases under subject-independent evaluation for all datasets, which is intuitive. These results confirm that cross-subject EEG classification is considerably more challenging than subject-dependent classification. However, the magnitude of the performance drop differs substantially between the two tasks. The MCPS EEG task for mTBI-ODD and MMIDB shows only a minor decrease (1.86\% and 0.60\% in F1 score, respectively), whereas the SCPS EEG task for ADFTD and PD-RS exhibits a significant decrease (49.77\% and 38.40\% in F1 score, respectively). This experiment alone does not explain the underlying cause of the performance drop or why it differs across MCPS and SCPS task types. However, this observation suggests that inter-subject variability alone may not fully account for the large performance drop observed in SCPS EEG classification, suggesting that additional factors may be involved.

\input{tables/dataset_differ}

\subsection{Controlling Dataset-Specific Differences}
\label{sub:dataset_differ}
To rule out the possibility that the observed difference in performance degradation is driven by dataset-specific factors, we conduct a controlled comparison using a dataset that contains both MCPS and SCPS tasks. We use the AOPD dataset, which includes both stimulus classification (standard vs. deviant, an MCPS task) and PD detection (HC vs. PD, an SCPS task), both of which are binary classification problems. We perform both subject-dependent and subject-independent evaluations for the two tasks while keeping all other settings identical. The results are presented in Table~\ref{tab:dataset_differ}.

The results show that even within the same dataset, the SCPS task still exhibits a much larger performance drop than the MCPS task when moving from subject-dependent to subject-independent evaluation. Specifically, the MCPS task shows only a 0.90\% decrease in F1 score, whereas the SCPS task shows a 30.66\% decrease in F1 score. This finding suggests that the difference in performance drop is related to the task formulation (MCPS vs. SCPS) rather than dataset differences. Since both tasks use the same dataset and the same train-test split, they should experience the same inter-subject variability. Therefore, the large performance drop observed in SCPS tasks must be attributed to factors beyond inter-subject variability.

\input{tables/subject_disc}

\subsection{Subject Discrimination Experiment}
\label{sub:sub_disc}
In SCPS EEG classification, each subject is assigned a single label, and all samples from the same subject share that label. Under subject-dependent evaluation, this setting may allow the model to use subject identity as a shortcut for prediction. Specifically, because all training samples from a given subject are associated with the same class label, the model may learn subject-specific patterns rather than task-relevant neural features. During testing, when samples from the same subjects are encountered again, the model may classify them based on their similarity to the corresponding training samples rather than on genuine task-related information. Consequently, the model can achieve high classification performance by exploiting subject identity instead of learning truly discriminative patterns.

This shortcut-learning mechanism is possible only if the EEG data contain sufficiently strong subject-identity information. To determine whether such information exists and can be captured by the model, we conduct a subject-discrimination experiment in which subject ID is treated as the classification label. High subject-discrimination performance would indicate that the model can effectively encode subject-specific features, which could subsequently be exploited as shortcuts in SCPS classification. The results are presented in Table~\ref{tab:sub_disc}.

The results show an extremely high F1 score of 98.59\% on the subject-discrimination task, even higher than its performance on the SCPS task under subject-dependent evaluation (98.35\%). Notably, subject discrimination is a 50-class classification problem (50 subjects), whereas the SCPS task of AOPD is a binary classification problem. These results provide strong evidence that EEG signals contain highly discriminative subject-specific information and that the model can effectively capture such information.

\input{tables/label_permutation}

\subsection{Label Permutation Experiments}
\label{sub:label_permutation}

To rigorously determine whether the model exploits subject-specific features as learning shortcuts, we conduct label permutation experiments on the AOPD dataset. The objective is to completely \textbf{eliminate genuine task-related information} while preserving the marginal label distribution, thereby isolating the model's reliance on subject identity. We employ task-specific permutation strategies (see Table~\ref{tab:toy_perm} for toy examples): 
\begin{itemize}
    \item \textbf{For the MCPS Task:} Because each subject is associated with multiple labels, the true task label varies within each subject and hence is assigned at the individual sample level. Accordingly, we apply permutation at the \textit{sample-level}. Shuffling labels across all individual samples completely decouples the EEG signals from the true task labels. This serves as a comparative baseline to verify that, without a static subject-to-label mapping, the model fundamentally requires true task correlations to achieve better-than-random performance.
    
    \item \textbf{For the SCPS Task:} Because each subject is associated with a fixed label, the true task label is inherently assigned at the subject level. Consequently, we apply permutation at the \textit{subject-level}. By shuffling labels across subjects, thereby assigning a single permuted label to all samples from a given subject, we intentionally destroy any genuine task-related information while preserving the 1:1 mapping between a subject and an arbitrary label. If the model relies on true task features, performance should collapse after permutation; conversely, if it relies on subject-specific shortcuts, performance will remain artificially high.
\end{itemize}

Theoretically, if the model learns only genuine task-relevant features, removing the true labels should degrade performance to the random level ($\approx$50\%) across all evaluation settings. Table~\ref{tab:label_permutation} reports the empirical results of these permutation tests. For the MCPS task, performance degrades to near-random levels under both evaluation schemes, with the F1 score dropping from 65.25\% to 49.77\% in the subject-dependent setting, and from 64.35\% to 48.92\% in the subject-independent setting. Because labels naturally vary within a single subject in the MCPS task, memorizing a subject identity alone provides no predictive power. These random level performances validate that subject identity cannot serve as shortcuts in MCPS tasks.

Conversely, the SCPS task demonstrates a severe vulnerability to shortcut learning. Under the subject-dependent setting, performance remains highly robust despite the complete removal of true task labels, with the F1 score experiencing a negligible decrease from 98.35\% to 98.15\%. However, under subject-independent evaluation, where subject overlap between the training and test sets is strictly prevented, performance collapses to random level (49.60\%). This massive discrepancy indicates that during subject-dependent SCPS evaluations, the model simply memorizes subject-specific EEG signatures and associates them with the available label. Once this subject overlap is removed in the subject-independent setting, the memorized mapping is rendered obsolete. This further highlights that the high performance achieved under subject-dependent evaluation for SCPS tasks is meaningless, as the model may learn little or nothing about task-relevant features.

%% file: tables/subject-dep-indep.tex
\begin{table}[h]
\centering
\def\arraystretch{1.0}
\caption{\textbf{Performance Drop under Subject-Independent Evaluation.} Comparison of classification performance for an MCPS and an SCPS EEG task under subject-dependent (Sub-Dep) and subject-independent (Sub-Indep) setups. 
}
\vspace{-2mm}
\label{tab:sub_dep_indep}
\resizebox{\columnwidth}{!}{%
\begin{tabular}{l|l |cc}
    \toprule
    
    \multicolumn{1}{l}{Datasets} & \multicolumn{1}{l|}{Setups} & \multicolumn{1}{c}{F1 Score} & \multicolumn{1}{c}{AUROC} \\ 
    \midrule

    \multirow{2}{*}{\begin{tabular}[c]{@{}l@{}}\;\makecell{mTBI-ODD \\ (MCPS)}\end{tabular}} 
    & Sub-Dep & 67.63\std{1.08} & 87.52\std{0.63}  \\
    & Sub-Indep & 65.77\std{2.11}\textcolor{red!70!black}{\scriptsize $\downarrow$1.86} & 85.90\std{1.23}\textcolor{red!70!black}{\scriptsize $\downarrow$1.62}  \\
    \cmidrule{1-4}
    \multirow{2}{*}{\begin{tabular}[c]{@{}l@{}}\;\makecell{MMIDB \\ (MCPS)}\end{tabular}} 
    & Sub-Dep & 65.98\std{1.87} & 72.53\std{1.83}  \\
    & Sub-Indep & 65.38\std{2.21}\textcolor{red!70!black}{\scriptsize $\downarrow$0.60} & 71.26\std{2.59}\textcolor{red!70!black}{\scriptsize $\downarrow$1.27}  \\
    \cmidrule{1-4}

    \multirow{2}{*}{\begin{tabular}[c|]{@{}l@{}}\;\makecell{\;ADFTD \\ \;(SCPS) }\end{tabular}} 
    & Sub-Dep & 97.66\std{0.46} & 99.85\std{0.05}  \\
    & Sub-Indep  & 47.89\std{3.93}\textcolor{red!70!black}{\scriptsize $\downarrow$49.77} & 67.46\std{4.07}\textcolor{red!70!black}{\scriptsize $\downarrow$32.39}  \\
    \cmidrule{1-4}
    \multirow{2}{*}{\begin{tabular}[c|]{@{}l@{}}\;\makecell{\;PD-RS \\ \;(SCPS) }\end{tabular}} 
    & Sub-Dep & 98.55\std{0.27} & 99.89\std{0.05}  \\
    & Sub-Indep  & 60.15\std{5.91}\textcolor{red!70!black}{\scriptsize $\downarrow$38.40} & 66.51\std{8.01}\textcolor{red!70!black}{\scriptsize $\downarrow$33.38}  \\

    \bottomrule 
\end{tabular}
}
\vspace{-2mm}
\end{table}

%% file: tables/dataset_differ.tex
\begin{table}[h]
\centering
\def\arraystretch{1.0}
\caption{\textbf{Controlling Dataset-Specific Differences.} Comparison of classification performance for an MCPS and an SCPS EEG task under subject-dependent (Sub-Dep) and subject-independent (Sub-Indep) setups using the same dataset. 
}
\vspace{-2mm}
\label{tab:dataset_differ}
\resizebox{\columnwidth}{!}{%
\begin{tabular}{l|l |cc}
    \toprule
    
    \multicolumn{1}{l}{Datasets} & \multicolumn{1}{l|}{Setups} & \multicolumn{1}{c}{F1 Score} & \multicolumn{1}{c}{AUROC} \\ 
    \midrule

    \multirow{2}{*}{\begin{tabular}[c]{@{}l@{}}\;\makecell{AOPD \\ (MCPS)}\end{tabular}} 
    & Sub-Dep & 65.25\std{1.24} & 72.02\std{1.64}  \\
    & Sub-Indep & 64.35\std{1.66}\textcolor{red!70!black}{\scriptsize $\downarrow$0.90} & 69.98\std{1.88}\textcolor{red!70!black}{\scriptsize $\downarrow$2.04}  \\
    \cmidrule{1-4}

    \multirow{2}{*}{\begin{tabular}[c|]{@{}l@{}}\;\makecell{AOPD \\ (SCPS) }\end{tabular}} 
    & Sub-Dep & 98.35\std{0.59} & 99.86\std{0.05}  \\
    & Sub-Indep & 67.69\std{9.23}\textcolor{red!70!black}{\scriptsize $\downarrow$30.66} & 74.15\std{11.95}\textcolor{red!70!black}{\scriptsize $\downarrow$25.71}  \\

    \bottomrule 
\end{tabular}
}
\vspace{-3mm}
\end{table}

%% file: tables/subject_disc.tex
\begin{table}[h]
\centering
\def\arraystretch{1.0}
\caption{\textbf{Subject Discrimination Experiment.} Perform subject-discrimination (Sub-Disc) by classifying subject ID. Subject-dependent (Sub-Dep) and subject-independent (Sub-Indep) results are also provided for comparison. 
}
\vspace{-2mm}
\label{tab:sub_disc}
\resizebox{\columnwidth}{!}{%
\begin{tabular}{l|l |cc}
    \toprule
    
    \multicolumn{1}{l}{Datasets} & \multicolumn{1}{l|}{Setups} & \multicolumn{1}{c}{F1 Score} & \multicolumn{1}{c}{AUROC} \\ 
    \midrule

    \multirow{2}{*}{\begin{tabular}[c]{@{}l@{}}\;\makecell{AOPD \\ (MCPS)}\end{tabular}} 
    & Sub-Dep & 65.25\std{1.24} & 72.02\std{1.64}  \\
    & Sub-Indep & 64.35\std{1.66}\textcolor{red!70!black}{\scriptsize $\downarrow$0.90} & 69.98\std{1.88}\textcolor{red!70!black}{\scriptsize $\downarrow$2.04}  \\
    \cmidrule{1-4}

    \multirow{2}{*}{\begin{tabular}[c|]{@{}l@{}}\;\makecell{AOPD \\ (SCPS) }\end{tabular}} 
    & Sub-Dep & 98.35\std{0.59} & 99.86\std{0.05}  \\
    & Sub-Indep & 67.69\std{9.23}\textcolor{red!70!black}{\scriptsize $\downarrow$30.66} & 74.15\std{11.95}\textcolor{red!70!black}{\scriptsize $\downarrow$25.71}  \\
    \cmidrule{1-4}

    \multirow{1}{*}{\begin{tabular}[c|]{@{}l@{}}\;\makecell{AOPD}\end{tabular}} 
    & Sub-Disc & 98.59\std{0.88} & 99.98\std{0.01}  \\

    \bottomrule 
\end{tabular}
}
\vspace{-3mm}
\end{table}

%% file: tables/label_permutation.tex
\begin{table}[h]
\centering
\caption{\textbf{Label Permutation Toy Examples.} Label permutation for MCPS and SCPS EEG tasks, where tuples stand for \textbf{(label, subject-id)}. Label permutation is used instead of random reassignment to preserve the original label distribution.}
\vspace{-2mm}
\label{tab:toy_perm}
\resizebox{\columnwidth}{!}{%
\begin{tabular}{cc|cc}
    \toprule
    \multicolumn{2}{c}{MCPS Label Permutation} &  \multicolumn{2}{c}{SCPS Label Permutation}  \\
    \midrule
    Original & Permutated & Original & Permutated \\
    \midrule
    (0, S1) & (1, S1) & (0, S1) & (1, S1) \\
    (1, S1) & (0, S1) & (0, S1) & (1, S1) \\
    (0, S1) & (0, S1) & (0, S1) & (1, S1) \\
    (1, S2) & (1, S2) & (1, S2) & (0, S2) \\
    (0, S2) & (1, S2) & (1, S2) & (0, S2) \\
    (1, S2) & (0, S2) & (1, S2) & (0, S2) \\
    \bottomrule
\end{tabular}
}
\vspace{-2mm}
\end{table}

\begin{table}[h]
\centering
\def\arraystretch{1.0}
\caption{\textbf{Label Permutation Experiments.} We permutate the labels for MCPS and SCPS tasks to prove the existence of shortcut learning using subject identity features. We use \underline{underline} to highlight random level performance.
}
\vspace{-2mm}
\label{tab:label_permutation}
\resizebox{\columnwidth}{!}{%
\begin{tabular}{l|l |cc}
    \toprule
    
    \multicolumn{1}{l}{Datasets} & \multicolumn{1}{l|}{Setups} & \multicolumn{1}{c}{F1 Score} & \multicolumn{1}{c}{AUROC} \\ 
    \midrule

    \multirow{4}{*}{\begin{tabular}[c]{@{}l@{}}\;\makecell{AOPD \\ (MCPS)}\end{tabular}} 
    & Sub-Dep & 65.25\std{1.24} & 72.02\std{1.64}  \\
    & Sub-Dep-Perm & \underline{49.77\std{1.85}} & \underline{50.47\std{2.27}}  \\
    \cmidrule{2-4}
    & Sub-Indep & 64.35\std{1.66} & 69.98\std{1.88}  \\
    & Sub-Indep-Perm & \underline{48.92\std{0.34}} & \underline{49.73\std{0.65}}  \\
    \cmidrule{1-4}

    \multirow{4}{*}{\begin{tabular}[c|]{@{}l@{}}\;\makecell{AOPD \\ (SCPS) }\end{tabular}} 
    & Sub-Dep & 98.35\std{0.59} & 99.86\std{0.05}  \\
    & Sub-Dep-Perm & 98.15\std{0.29} & 99.79\std{0.10}  \\
    \cmidrule{2-4}
    & Sub-Indep & 67.69\std{9.23} & 74.15\std{11.95}  \\
    & Sub-Indep-Perm & \underline{49.60\std{8.84}} & \underline{50.04\std{13.82}}  \\

    \bottomrule 
\end{tabular}
}
\vspace{-3mm}
\end{table}

%% file: 040discussion.tex
\begin{figure*}[t]
    \centering    \includegraphics[width=1.0\linewidth]{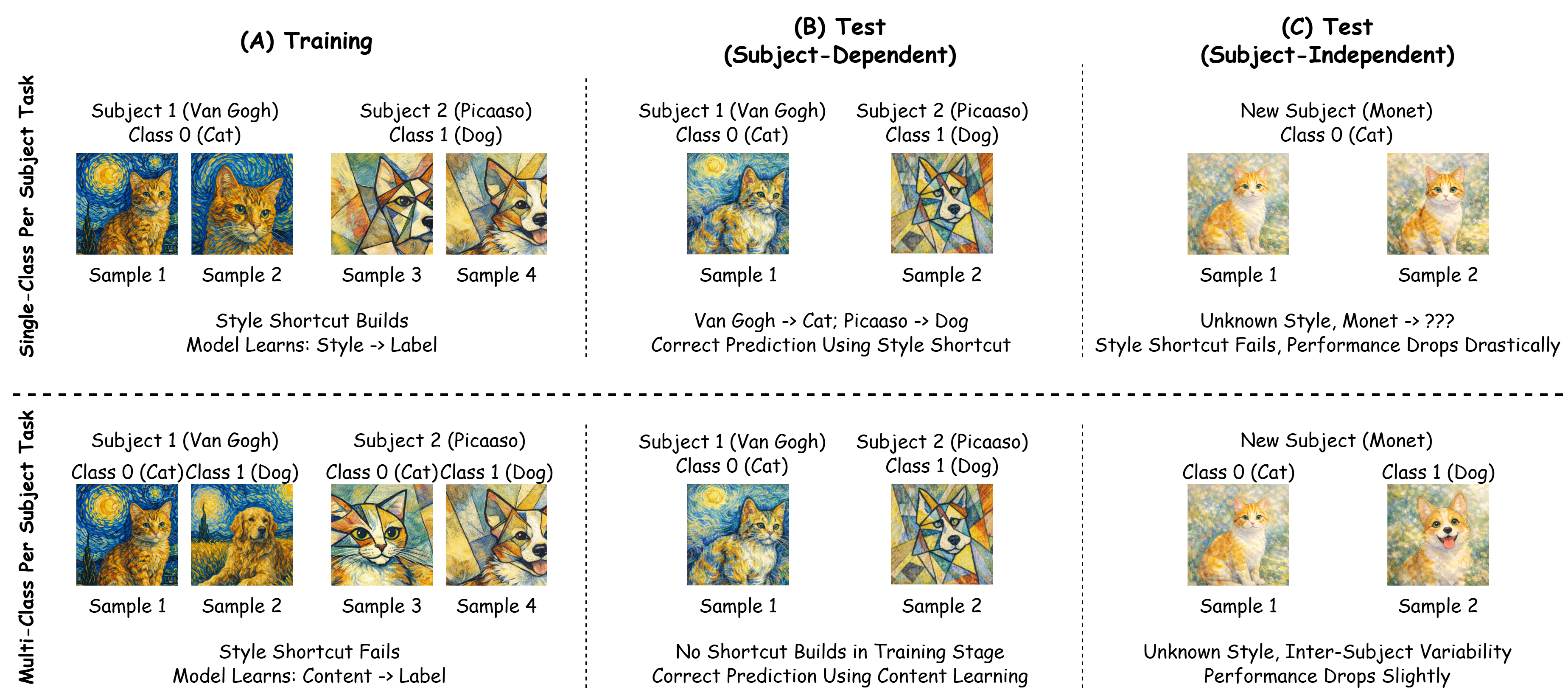}
    \caption{ \textbf{Shortcut Learning Illustration Using an Image Analogy.} Subject Identity = Image Style (e.g., Van Gogh, Picasso); Task-relevant Features = Image Content (e.g., cat, dog). In SCPS tasks (top row), each subject is associated with a single class, allowing the model to learn a shortcut by mapping style to label. This leads to strong performance under subject-dependent evaluation but poor generalization to unseen subjects under subject-independent evaluation. In contrast, in MCPS tasks (bottom row), each subject contains multiple classes, preventing the model from relying on style-based shortcuts and encouraging learning of content-related features, resulting in better generalization.
    }
    \label{fig:shortcut_learning}
    \vspace{-5mm}
\end{figure*}

\subsection{Global Analysis}
\label{sub:global_analysis}
Our experiments suggest that performance degradation in cross-subject EEG classification is mainly caused by two factors: \textbf{inter-subject variability} and \textbf{shortcut learning}. Inter-subject variability affects both MCPS and SCPS tasks by inducing distribution shifts between training and test subjects. In contrast, shortcut learning primarily affects SCPS tasks, in which subject identity is strongly correlated with class labels. Subject-independent evaluation prevents models from exploiting subject-identity shortcuts at test time, but it does not prevent them from learning such shortcuts during training. Consequently, subject-independent evaluation often leads to a substantial performance drop, particularly in SCPS tasks. These findings highlight the critical role of evaluation protocols in EEG classification research and suggest that subject-dependent evaluation may substantially overestimate model performance on SCPS tasks. Moreover, researchers developing EEG cross-subject domain adaptation methods should be aware of the distinct challenges posed by different tasks, as methods designed solely to mitigate inter-subject variability may remain ineffective against shortcut learning.

\subsection{Image Analogy of Shortcut Learning}
\label{sub:shortcut_learning}
To provide a more intuitive understanding of how shortcut learning arises in SCPS EEG classification, we present an analogy based on image classification in Figure~\ref{fig:shortcut_learning}. In this analogy, subject identity corresponds to image style (e.g., Van Gogh or Picasso), whereas the task-relevant label corresponds to image content (e.g., a cat or a dog). For the SCPS task shown in the top row, each subject is associated with only one class (e.g., Van Gogh always painted cats; Picasso always painted dogs). During training (A), samples from the same subject share both style and label, allowing the model to associate style with the label (e.g., Van Gogh $\rightarrow$ cat, Picasso $\rightarrow$ dog). As a result, a style-based shortcut emerges, and the model learns to map style directly to the label rather than learning content-related features. Under subject-dependent testing (B), the same subjects appear in both the training and testing sets. Consequently, the model can still classify samples correctly by recognizing their style, even without understanding the underlying content. However, under subject-independent testing (C), new subjects correspond to unseen styles (e.g., Monet). Because the learned shortcut no longer holds, the model fails to generalize, leading to a substantial drop in performance. In contrast, for the MCPS task shown in the bottom row, each subject contains samples from multiple classes (e.g., Van Gogh painted both cats and dogs). During training (A), style is no longer predictive of the label, so the model cannot rely on a style-based shortcut. Instead, it is forced to learn content-related features. As a result, performance remains relatively stable under subject-independent (C) testing on unseen subjects, with only a moderate decline caused by inter-subject variability. This illustration highlights that shortcut learning arises when subject identity is strongly correlated with labels, and explains why SCPS EEG classification is particularly vulnerable to performance degradation under subject-independent evaluation.

\subsection{Future Work}
\label{sub:future_work}
First, as discussed earlier, the subject-independent setup prevents shortcut learning at test time but does not prevent the model from learning subject-identity features during training. Future work should focus on developing methods to disentangle subject-identity features from task-related features during training. Second, beyond algorithm-level solutions to alleviate shortcut learning in SCPS EEG tasks, data-level solutions may also be feasible. For example, increasing the number of subjects may reduce the impact of subject identity shortcuts. In addition, large-scale pre-trained foundation models trained on diverse datasets may help alleviate this issue by learning more generalizable EEG representations. These directions are worth exploring in future work. Third, in this paper, we broadly define shortcut features as subject identity features. However, it remains unclear what these features actually are. EEG data are often collected across multiple sessions and trials on different days, sometimes with long time intervals between recordings. An interesting question is whether these subject identity features persist over time or change across sessions. More experiments are needed to investigate this issue.

%% file: 050conclusion.tex
In this paper, we systematically investigated the causes of performance degradation in cross-subject EEG classification through a series of controlled experiments on both MCPS and SCPS EEG tasks. Our results and analysis show that the performance drop observed when moving from subject-dependent to subject-independent evaluation cannot be explained by inter-subject variability alone. We demonstrate that performance degradation in cross-subject EEG classification is mainly caused by two different factors. In MCPS EEG tasks, performance degradation is primarily due to inter-subject variability, leading to domain shift between training and test subjects. For SCPS EEG tasks, however, performance degradation is largely caused by shortcut learning based on subject identity features. In subject-dependent evaluation, models can exploit subject identity information as a shortcut for classification when subject identity is strongly correlated with labels. When moving to subject-independent evaluation, subject identity shortcuts are no longer available, leading to a significant performance drop. Overall, this work shows that performance degradation in cross-subject EEG classification arises from distinct mechanisms, depending on the task formulation, highlighting the importance of proper evaluation protocols and inspiring future work on method design in EEG classification research.